# Understanding and controlling hexagonal patterns of wrinkles in neo-Hookean elastic bilayer structures

Teng Zhang[a,b]*

[a] Department of Mechanical and Aerospace Engineering, Syracuse University, Syracuse, NY 13244, United States. Email: tzhang48@syr.edu; Tel: +1 315-443-2969.

[b] BioInspired Syracuse: Institute for Material and Living Systems, Syracuse University, Syracuse, NY 13244, United States.

**Abstract**

A controlled surface wrinkling pattern has been widely used in diverse applications, such as stretchable electronics, smart windows, and haptics. Here, we focus on hexagonal wrinkling patterns because of their great potentials in realizing anisotropic and tunable friction and serving as a dynamical template for making non-flat thin films through self-assembling processes. We employ large-scale finite element simulations of a bilayer neo-Hookean solid (e.g., a film bonded on a substrate) to explore mechanical principles that govern the formation of hexagonal wrinkling patterns and strategies for making nearly perfect hexagonal patterns. In our model, the wrinkling instabilities are driven by the confined film expansion. Our results indicate robust hexagonal patterns exist at a relatively small modulus mismatch (on the order of 10) between the film and substrate. Besides, the film expansion should not exceed the onset of wrinkling value too much to avoid post-buckling patterns. By harnessing the imperfection insensitivity of one-dimension sinusoidal wrinkles, we apply a sequential loading to the bilayer structure to produce the nearly perfect hexagonal patterns. Lastly, we discuss the connection between the simple bilayer model and the gradient structures commonly existed in experiments.

# 1. Introduction

Surface wrinkles play critical roles in achieving various tunable mechanical and physical properties, such as adhesion [Chan et al., 2008; Lin et al., 2008], friction [Suzuki and Ohzono, 2016; Suzuki et al., 2014; Yuan et al., 2019], and wetting [Khare et al. 2009; Chunget al., 2007; Zhang et al., 2012; Wu et al., 2019; Zhang et al., 2020] by dynamically changing the surface morphologies and thus enable new devices for diverse engineering applications, such as stretchable electronics [Rogers et al., 2010], smart windows [Kim et al., 2013; Jiang et al., 2018; Kim et al., 2018; Jiang et al., 2019], and haptics [Skedung et al., 2013]. In the last two decades, significant effort has been devoted to uncovering the fundamental principles that govern the formation and evolution of the wrinkles and exploring efficient fabrication techniques [Genzer and Groenewold, 2006; Yang et al., 2010; Li et al., 2012; Wang and Zhao, 2016]. It has been well documented that controlled wrinkles can be formed by compressing a system with a stiff film bonded on a soft substrate or similar system with a gradient modulus. Especially for one-dimensional (1D) wrinkles, the wrinkle wavelength, critical strain, and post-buckling behaviors (e.g., period doubling, folding, and ridge) can be well understood in the framework of finite deformation elasticity and observed in experiments [Jiang et al., 2007; Song et al., 2008; Dervaux and Ben Amar, 2011; Li, et al., 2011; Mei et al., 2011; Brau et al., 2011; Sun et al., 2012; Cao and Hutchinson, 2012; Zang et al., 2012; Ni et al., 2012; Diab et al., 2013; Wang and Zhao, 2014; Zhao et al., 2015; Jin et al., 2015; Budday et al., 2015; Zhao et al., 2016; Auguste et al., 2017; Zhang, 2017; Jia et al., 2018; Sui et al., 2019]. Substantial progress has also been made for the two-dimensional (2D) wrinkles [Bowden et al., 1998; Chen and Hutchinson, 2004; Huang et al., 2005; Audoly and Boudaoud, 2008a, 2008b, 2008c; Guvendiren et al., 2009; Li, Jia, et al., 2011; Cai et al. 2011; Breid and Crosby, 2013;

Ciarletta et al., 2014; Xu et al., 2014; Terwagne et al., 2014; Tallinen et al., 2014; Huang et al., 2016; Zhang et al., 2019; Zhao et al., 2020; Xu et al., 2020]. For example, Guvendiren et al. [2009] investigated the swelling induced wrinkles in a gradient structure in poly (hydroxyethyl methacrylate) (PHEMA) hydrogels created by manipulating the UV light curing process. Various wrinkling patterns (e.g., hexagon and peanut) can be formed by varying the density of crosslinkers (Fig. 1a). Cai et al., [2011] applied an ultraviolet-ozone (UVO) oxidation process to a PDMS structure to generate stiff surface layers and triggered the wrinkling instabilities by the absorption of ethanol vapor. Depending on the UVO treatment times, they observed hexagonal (short expose time) and herringbone patterns (long expose time), as shown in Fig. 1b. Further, hexagonal and labyrinth patterns are found in a spherical bilayer structure (Fig. 1c) where the release of the pre-strain in the inner layer leads to compression of the outer thin films and eventually the formation of wrinkles [Terwagne et al., 2014; Brojan et al., 2015; Stoop et al., 2015].

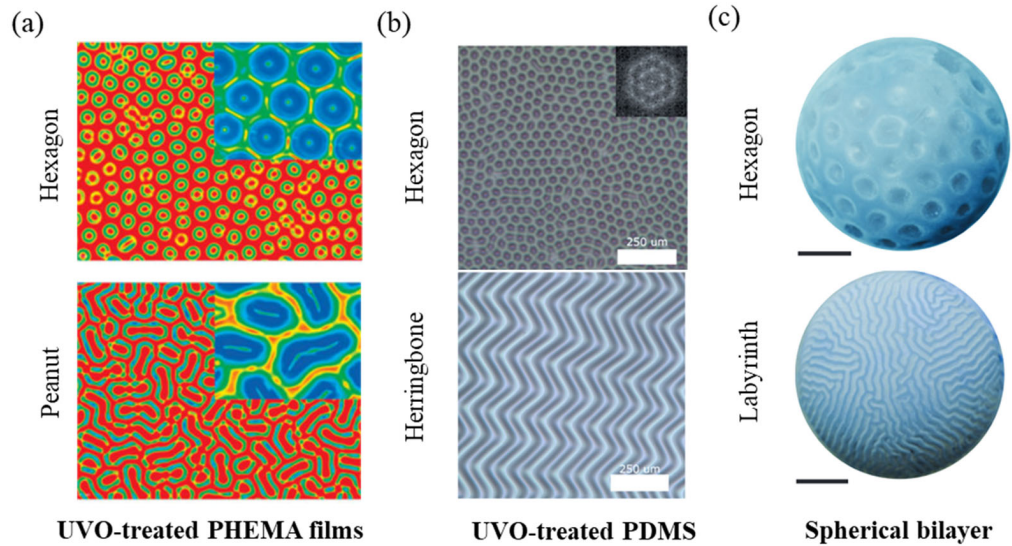

**Figure 1.** Wrinkling patterns in experiments. (a) Hexagon and peanut patterns are observed in the swelled PHEMA films treated by UVO, adapted from Guvendiren et al. [2009]. (b) Hexagon and herringbone patterns in UVO-treated PMDS due to swelling, adapted from Cai et al. [2011]. (c) Hexagon and Labyrinth patterns on spherical bilayers with a top layer bonded on a pre-stretched inner layer adapted from Stoop et al. [2015].

Among all these patterns, hexagonal patterns have attracted a lot of attention because well-controlled hexagons can create a new avenue to tailor the surface for certain functions, such as anisotropic and tunable friction and a dynamical template for making non-flat thin films through self-assembling processes. Terwagne et al. [2014], fabricated spherical surfaces with hexagonal wrinkles, the morphable surfaces with dimples were also employed to dynamically control aerodynamic drag. Further, Brojan et al. [2015], proposed a wrinkling crystallography concept to describe the dimple patterns they found on the spherical surfaces, which opens the possibility of bridging the wrinkling patterns with crystal structures. To quantify the energy levels of different wrinkling patterns, Cheewaruangroj and Biggins [2019] conducted high order perturbation analysis and finite element (FE) simulations of the stripe (1D), square, and hexagon in a flat bilayer structure. Their results showed that the hexagonal pattern, more specifically the positive hexagon (dents or dimples), has the lowest energy among the wrinkling modes due to the break of invasion symmetry of the hexagons. Their work provides a reasonable explanation of the formation of the hexagons but cannot fully explain the formation of herringbone or labyrinth pattern and peanut patterns. In addition, most reported wrinkling patterns in the experiments are not perfect hexagons and contains pentagons and heptagons, which can be viewed as defects (e.g., disclinations, dislocations, and grain boundaries) if we treat hexagons as perfect crystal lattices. Very recently, hexagonal patterns down to micrometers were also reported in nanoscale strained $MoS_2$ film on a sapphire substrate [Ren et al., 2019].

The current work is to uncover the mechanical principles that govern the formation of hexagonal wrinkling patterns and strategies for making nearly perfect hexagons. We adopt large-scale FE simulations to address this highly nonlinear problem. To simplify this highly nonlinear problem, we focus on a bilayer structure, made of neo-Hookean solids. The rest paper is organized as follows. Chapter 2 discusses the simulation model. The main results are reported in Chapter 3, including the effect of modulus mismatch on the pattern formation, post-buckling behaviors, and the control of hexagons with a sequential loading. Finally, some concluding remarks are given in Chapter 4.

**2. The simulation model**

We focus on a simple bilayer structure with a thin and stiff film bonded on a large and soft substrate, as shown in Fig. 2a. In the following analysis, we set film thickness $h = 1$ and normalize the rest of lengths with respect to $h$. To capture a large number of wrinkles and their nonlinear interactions, we choose the structure sizes are all much larger than the thickness (e.g., $\bar{L}_x = L_x/h > 200$, $\bar{L}_y = L_y/h > 200$, and $\bar{L}_z = L_z/h > 200$). We assign isotropic expansion to the thin film to mimic the volumetric change during swelling. The side surfaces are confined to generate compressive stress in the film (Fig. 2a). When the compressive stress exceeds a critical value, wrinkling instabilities occur.

Both film and substrate are described by neo-Hookean model, whose strain energy density can be expressed as

$$U = \frac{1}{2}\mu(\bar{I}_1 - 3) + \frac{1}{2}K(J^{el} - 1)^2, \qquad (1)$$

where $\mu$ and $K$ are shear and buckle modulus, respectively, $\bar{I}_1$ is the first deviatoric strain invariant defined as $\bar{I}_1 = \bar{\lambda}_1^2 + \bar{\lambda}_2^2 + \bar{\lambda}_3^2 = J^{-\frac{2}{3}}(\lambda_1^2 + \lambda_2^2 + \lambda_3^2)$ and $\lambda_i$ are the principal stretches and the total volume ratio $J = \det(\boldsymbol{F}) = \lambda_1\lambda_2\lambda_3$, $\boldsymbol{F}$ is the deformation gradient tensor, and $J^{el}$ is the elastic volume ratio. The isotropic expansion in the film can be implemented in the model by defining $J^{el}$ as

$$J^{el} = \frac{J}{(1+\varepsilon^p)^3}, \qquad (2)$$

where the $\varepsilon^p$ is a controlling strain-like parameter for the expansion. The nearly incompressible neo-Hookean material is achieved by setting $K/\mu = 20$, which corresponds to a Poisson's ratio as 0.475.

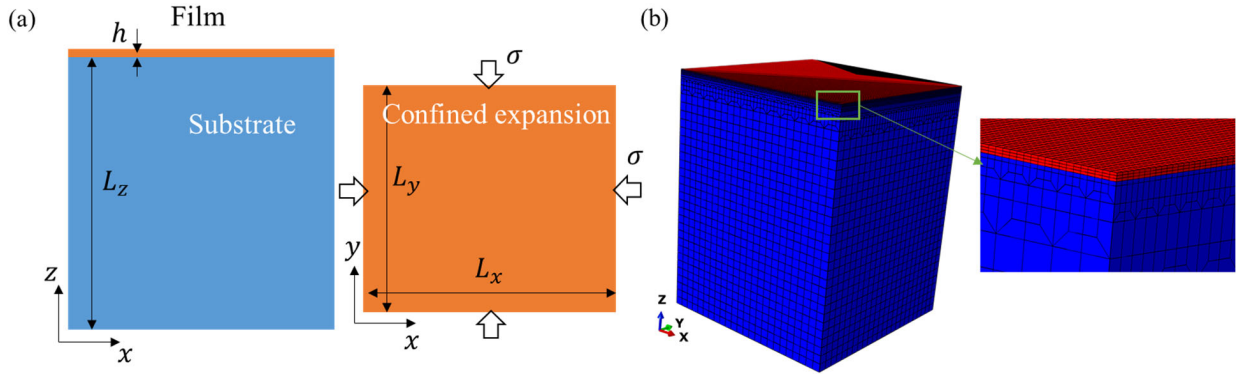

**Figure 2.** Simulation model set up. (a) Schematic of the bilayer structure with a thin film (yellow) and thick substrate (blue). The confined film expansion generates compressive stress in the film to trigger the formation of wrinkles. (b). FEM mesh for a representative bilayer with $\bar{L}_x = \bar{L}_y = 400$ and $\bar{L}_z = 500$. Here we set $h = 1$.

All the FE simulations are carried out with ABAQUS/Standard. We choose the 20-node quadratic brick (hexahedral) element (C3D20) to simulate the large nonlinear deformation of the wrinkles. To ensure accuracy, we keep at least three elements along the film thickness direction and make sure each wrinkle unit is discretized by more than ten elements. A gradient mesh is adopted to smoothly transit from fine mesh at the top surfaces to coarse mesh at the bottom mesh (Fig. 2b). The total FE mesh can still reach 1 million even with the help of this technique. Therefore, we employ the iterative solver for the linear equations inside the Newton solver to alleviate the high demand of computational memory. The film expansion is implemented using the "Expansion" function in ABAQUS. The side and bottom surfaces are confined by assigning the displacement along the surface normal direction as zero while the top surface is traction free. We further add a Gaussian white noise on the z coordinates of nodes on the top surface to trigger the instabilities. The mean magnitude of the noise is 1% of film thickness ($h$), which can trigger instabilities while does not significantly alter the critical strain for onset wrinkling (see supplemental materials for details). During simulations, a small artificial damping term is also applied to the equilibrium equation to help the convergence.

## 3. Results

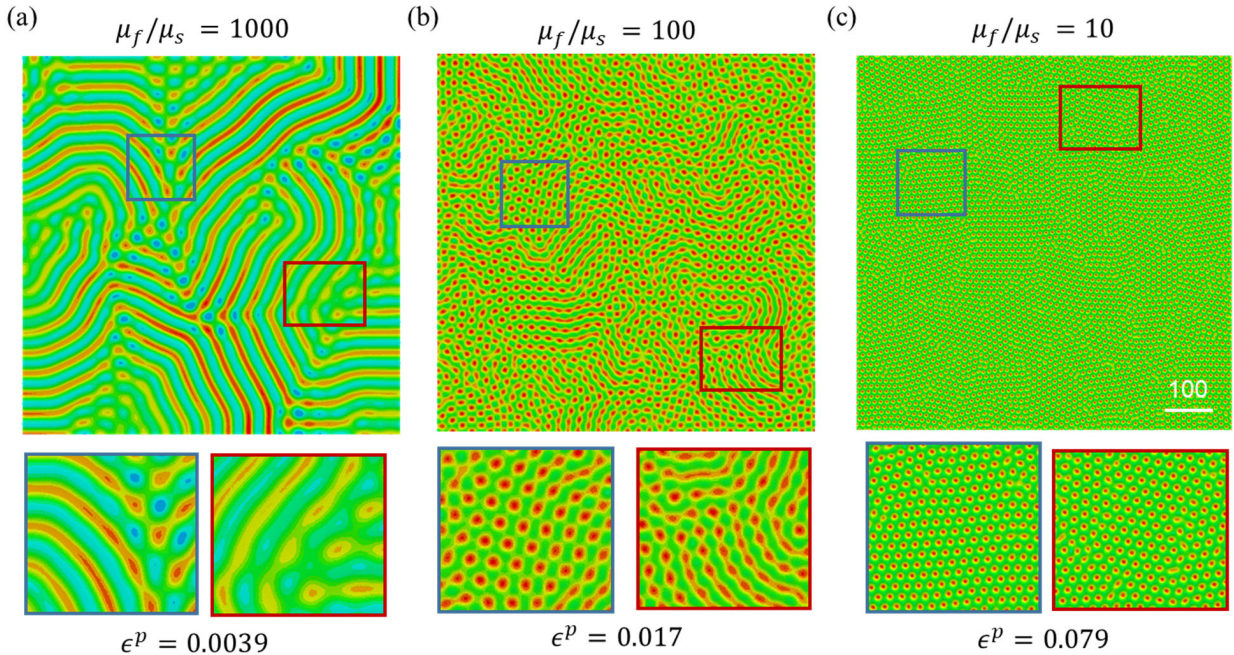

**Figure 3.** The effect of modulus mismatch on wrinkling patterns. (a) Labyrinth-like patterns are observed for very large modulus mismatch ($\mu_f/\mu_s = 1000$). (b) Mixed patterns (hexagons and square) are observed for intermediate modulus mismatch ($\mu_f/\mu_s = 100$). Labyrinth-like patterns start form as well. (c) Hexagon patterns are observed for relatively small modulus mismatch ($\mu_f/\mu_s = 10$). The bottom ones are zoom-in images to show the patterns more clearly.

We first examine the effect modulus mismatch on the wrinkling patterns in the bilayer structure. This is motivated by the fact that herringbone and labyrinth patterns have been widely reported in systems with a very stiff film on a soft substrate, such as metal film on polymeric materials [Bowden et al., 1998] and glassy film on elastomers [Cai et al., 2011]. We simulate a large structure with $\bar{L}_x = 800$ and $\bar{L}_y = 800$ to accommodate enough wrinkling units even for a very stiff film (e.g., $\mu_f/\mu_s = 1000$). Three representative modulus mismatch ratios, including $\mu_f/\mu_s = 1000, 100, 10$ are adopted here. For all the simulations, we keep the geometries including the perturbed top surface nodes and the modulus of the substrate unchanged, and only vary the film modulus and the value of the expansion. In all the figures, the color represents the von Mises stress and is only used to show the wrinkling patterns. Since the absolute stress value is not critical information, the color bars are omitted in all the figures for the sake of simplicity.

For the very large modulus mismatch ($\mu_f/\mu_s = 1000$), we find that a random labyrinth-like pattern is formed at expansion strain around 0.004 (Fig. 3a), without a clear indication of the initial formation of hexagons or squares. When we reduce the modulus mismatch to 100, a mixed hexagon and square can be noticed for a very short range of film expansion strain (Fig. 3b). For the relatively small modulus mismatch (e.g., $\mu_f/\mu_s = 10$), a dominated hexagonal pattern can be found with defective modes of heptagon and pentagon (Fig. 3c), which may be attributed to random nucleation of the patterns on perturbed surfaces. These results clearly show hexagons tend to form for relatively small modulus mismatch, which is indeed consistent with the energy analysis of different patterns in a unit cell [Cheewaruangroj and Biggins, 2019]. For the more complicated patterns exhibited in the large modulus mismatch cases, we speculate the different wrinkling patterns may emerge almost at the same time due to geometric imperfections. The interaction of these modes further leads to mixed patterns or post-buckling patterns without showing a clear onset wrinkling pattern. To obtain robust hexagonal wrinkling patterns, we need to eliminate this imperfection sensitivity and thus design a bilayer with a relatively small modulus mismatch.

We next turn to the post-buckling behavior of bilayer structure under larger film expansion. The simulated structure has an in-plane dimension as with $\bar{L}_x = \bar{L}_y = 240$ and modulus mismatch as $\mu_f/\mu_s = 5$. We choose this modulus mismatch because its post-buckling behavior resembles the nonlinear pattern evolutions of UVO-treated PDMS during swelling [Cai et al., 2011]. As shown in Fig. 4a, we find hexagon dominated patterns at a film expansion strain as 0.13. When the film expansion strain is increased to 0.17, the hexagons evolve into segment labyrinth patterns. Further, the simulation snapshots during the evolution are in great agreement with the experimental images of a UVO-treated PDMS during swelling (Fig. 4b). More interestingly, our simulations indicate the localizations first start at the heptagons (purple polygon in Fig. 4b), which is also consistent with the experimental images. This analysis of the post-buckling shows that the film expansion strain in the film should be controlled in a certain range to maintain the hexagonal patterns.

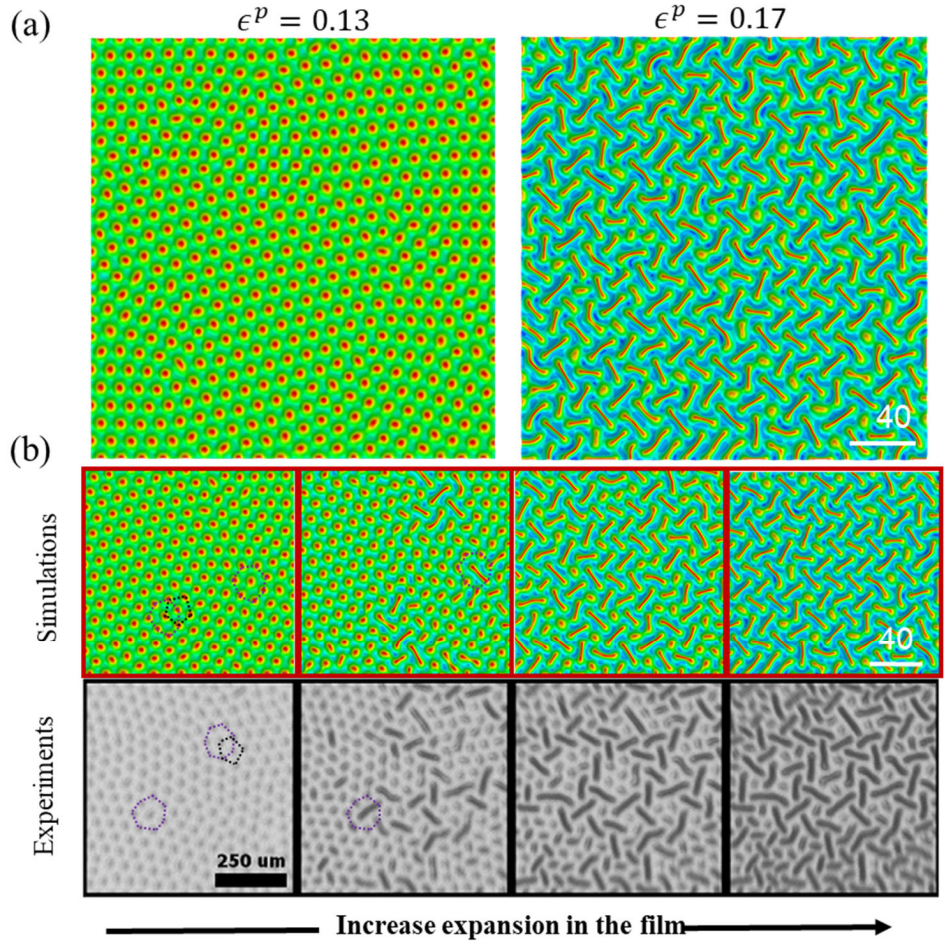

**Figure 4.** The formation and evolution of hexagonal patterns with increased film expansion with $\mu_f/\mu_s = 5$ and $\bar{L}_x = \bar{L}_y = 240$. (a) Hexagonal patterns at expansion strain of 0.13. (b) Segment labyrinth patterns at expansion strain of 0.17. (c) Comparison of the pattern evolution between simulations and experiments as the increase of film expansion. The experimental results are adapted from Cai et al. [2011].

Lastly, we explore the means of designing perfect hexagonal patterns, with as few as possible heptagons and pentagons, which are defined as defects here. As shown in our simulations and reported experiments, defects widely exist in the wrinkling patterns. This is likely due to various imperfections in the simulations (e.g., the perturbed surface nodal coordinates) and experiments (e.g., roughness and modulus heterogeneous). These imperfections are usually hard to be fully removed. In simulations, imperfections are needed to trigger the wrinkling patterns. Therefore, it is inevitable to deal with imperfections both in experiments and simulations. By noticing that the 1D sinusoidal wrinkles are not sensitive to the imperfections due to the strong anisotropy in the configuration and stress state, we create a sequential loading strategy to control the patterns. In step 1, we stretch the structure along the y direction during the expansion to release the confinement in that direction. This leads to sinusoidal wrinkles perpendicular to y direction. In step 2, we compress the structure along the y direction to its original length and re-create an equal-biaxial stress state in the structure. This gives a hexagon dominated patterns (Fig. 5a). On the contrary, many defects are formed when the whole structure under fully confinement along x and y directions (Fig. 5b), which maintains the equal-biaxial stress state all the time. We further construct the Voronoi diagrams by viewing each local stress maximum point as a particle (a polygon center) and present a detailed analysis of the hexagon, heptagon, and pentagon in Fig. 5c. The Voronoi diagrams confirm the visual assessment that the sequential loading structure only has two defects (pentagon-heptagon pair) while the one-step loading structure has more than 80 defects (42 pentagons and 41 heptagons). The imperfection insensitive hexagonal patterns in the two-step loading can be understood as follows. In the first step of the two-step loading, a long strip pattern is energetically preferred, whose number of pattern (n) is much smaller than the final hexagons ($n^2$). Therefore, the possible nucleation sites in the strip patterns are much less than the hexagons. In addition, the defects in strip patterns can only be dislocations (i.e., inserting one strip from one side), while many defects in the hexagons can be dislocation dipoles, whose nucleation energy barriers are generally much smaller than dislocations. Starting from almost uniform strip pattern, we can achieve a nearly perfect hexagonal pattern. It should be noted that sequential loading has also been employed to design perfect herringbone patterns [Yin et al., 2012]. Therefore, the sequential loading method is a good strategy to make perfect hexagon patterns by harnessing the imperfection insensitivity of the intermediate 1D wrinkles.

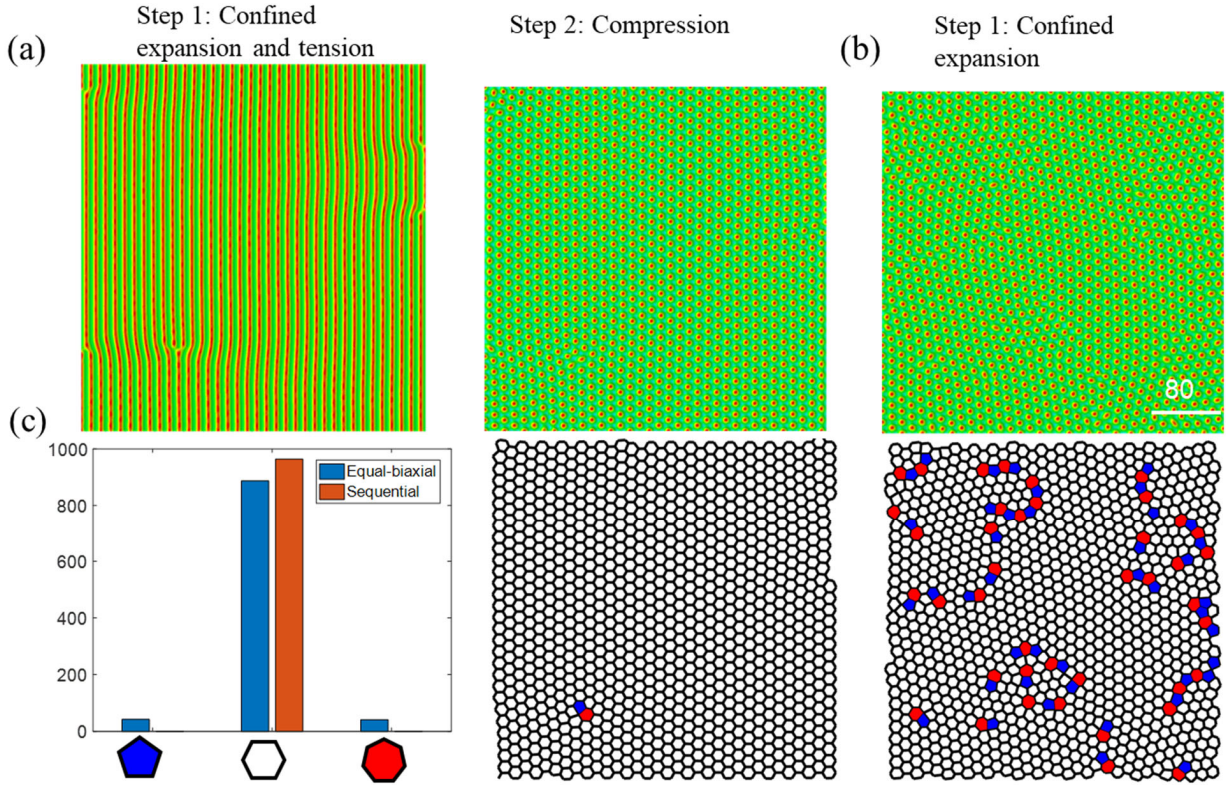

**Figure 5.** Nearly perfect hexagon pattens obtained from sequential loading. (a) Only one dislocation can be found in the hexagon patterns for the two-step loading case. (b) More defects in the hexagon patterns for the one-step loading case. (c) Statistics of the hexagons, heptagons, and pentagons in (a) and (b). The expansion $\epsilon^p$ in (a) and (b) is 0.08.

## 4. Discussion and conclusion

Our study systematically examines the principles of formation and control of hexagonal wrinkling patterns in neo-Hookean bilayer structures. It should be noticed that many experimentally reported studies are based on the gradient structures [Guvendiren et al., 2009; Glatz, B.A. and Fery, 2019], which is largely due to the diffusion-controlled process of creating a stiff top layer, such as UVO treatment. Although our model is much more simplified than the real gradient structures, the great agreement between the simulated pattern formation and evolution and experimental reports indicate that our findings can also be appliable to these structures. In fact, the gradient modulus distribution can help to explain why hexagonal patterns can appear for elastomer with a glassy top surface (the modulus mismatch can easily reach 1000). If we take the gradient decayed modulus into account and construct an effective bilayer, the effective film modulus will be much smaller than the top surface glassy layer. This leads to a relatively small modulus mismatch and thus provides a plausible explanation of the formation of robust hexagon patterns. The search for an effective bilayer structure for a given gradient structure is a very interesting research topic and will be studied in future work. Another simplification of our current work is to mimic the swelling deformation with a simple volumetric expansion. It is well known that the gel-like materials will change their elastic modulus during swelling [Rubinstein et al., 1996; Bosnjak et al., 2020]. This can further alter the critical strain for wrinkling formation and wrinkling wavelength as well as post-buckling behavior. Therefore, much more work is still needed to incorporate the coupling between swelling and mechanical deformation to predict the wrinkling pattern formation and evolution more accurately in polymer gels. This is expected to provide another dimension to tune and design the wrinkling patterns by harnessing the elastic modulus change during swelling. On spherical surfaces, it has been shown that more defects appear than the topological constraint [Brojan et al., 2015]. It will also be very interesting to examine the effect of sequential loading on the defect distributions of wrinkling on curved surfaces, where an anisotropic deformation mode first trigger strip wrinkling patterns.

In summary, we performed large-scale FE simulations to understand the mechanical principles that govern the formation of hexagonal wrinkling patterns and strategies for making nearly perfect hexagon patterns. To simplify the analysis, we adopted neo-Hookean bilayer structures, where the film is under expansion to drive the formation of wrinkles. We found that robust hexagons exist at a relatively small ($\mu_f/\mu_s \sim 10$) modulus mismatch ratio between film and substrate. This is likely due to the imperfection sensitivity of various wrinkling patterns at large modulus mismatch. Because post-buckling (e.g., peanut patterns) can happen at larger film expansion, one should control the expansion in a certain range to keep the hexagons. We further applied a sequential loading strategy to the bilayer structure to create nearly perfect hexagons. The current finding of the control hexagonal patterns can provide a rotational design principle for making multifunctional surfaces with perfect and tunable hexagonal topographies.

**Acknowledgements**

TZ thanks the support from Syracuse University. Simulations were performed at the Triton Shared Computing Cluster at the San Diego Supercomputer Center, the Comet cluster (Award no.TG-MSS170004 to TZ) in the Extreme Science and Engineering Discovery Environment (XSEDE) and the Academic Virtual Hosting Environment (AVHE) at Syracuse University.